\documentstyle[multicol,prl,aps,epsf,epsfig]{revtex}

\begin{document}

\title{Understanding Quantum Superarrivals using the Bohmian model}

\author{ Md. Manirul Ali\footnote{mani@bose.res.in}$^1$, 
A. S. Majumdar\footnote{archan@bose.res.in}$^1$, and
Dipankar Home\footnote{dhom@boseinst.ac.in}$^2$}

\address{$^1$S. N. Bose National Centre for Basic Sciences,
Block JD,
Sector III, Salt Lake, Calcutta 700098, India}

\address{$^2$Department of Physics, Bose Institute, Calcutta
700009, India} 

\maketitle

\begin{abstract}

We investigate the orgin of ``quantum superarrivals'' in the reflection 
and transmission probabilities of a Gaussian wave packet  for a 
rectangular potential
barrier while it is perturbed by either reducing or increasing its height. 
There exists a 
finite time interval during which the probability of reflection is {\it larger}
(superarrivals) while the barrier is {\it lowered} compared to the 
unperturbed case. Similarly, during a certain interval of time, the probability of 
transmission while the
barrier is {\it raised} {\it exceeds} that for free propagation. We compute
{\it particle trajectories} using the Bohmian model of quantum mechanics 
in order to understand {\it how} this phenomenon of superarrivals occurs.

PACS number(s): 03.65.Bz

\end{abstract}

\begin{multicols}{2}

A number of interesting investigations have been reported on
wave-packet dynamics\cite{greenberger} 
including, in particular, recent studies on issues such as the 
observation of revivals of
wave packets\cite{venu}. Of late, 
we had pointed out a hitherto unexplored effect\cite{bandyo} considering
the time dependent reflection probability of a Gaussian wave packet 
reflected from a perturbed potential barrier. By reducing the height of the
barrier to zero in a short span of time during which there is a significant
overlap of it with the wave packet, we observed that the reflection probability
is {\it larger} compared to the case of reflection from a static barrier 
for a small but finite interval of time. This phenomenon is what we have called
``Quantum Superarrivals''. The speed with which the effect due to 
reducing the barrier 
height propagates across the wavefunction was noticed to be depending on the
rate at which the barrier height is reduced. We also found  the 
magnitude of superarrivals to be proportional to the rate of reduction
of the potential barrier. We argued that superarrivals occur because of the 
``objective reality'' of a wave function acting as a ``field'' which mediates
across it  the propagation of a physical disturbance, {\it viz}. perturbation 
of the potential barrier.

The aim of this paper is to further {\it generalize} the phenomenon  of superarrivals
and also to understand {\it how} superarrivals occur. 
We begin  by first showing that superarrivals also
indeed occur in the transmission probability when the barrier height is raised
from zero to some value (this is {\it complementary} to the superarrival 
phenomenon occuring for the reflected wave packet). We then compute particle 
trajectories using the Bohm model. We derive a {\it quantitative} {\it estimate} of
the magnitude of superarrivals using the Bohmian trajectories. We show that
it is possible to obtain a deeper insight into the nature of superarrivals
using such computed trajectories of individual particles. We illustrate this by
considering the case of a wave packet which is  reflected from the perturbed barrier.
Similar analysis can be done for the transmitted wave packet.

Let us first briefly recapitulate the essential features of quantum 
superarrivals. Consider a Gaussian wave packet peaked at $x_0$ with
half width $\sigma$. It moves to the right and strikes a potential barrier
of width $w$ centred at a point $x_c$. A detector  placed at a 
point $x'$ far left of $x_0$ 
measures the time-dependent reflection probability
by counting the reflected particles arriving there up to various instants
for both the case of a static barrier, and also when the barrier is 
perturbed by reducing its height to zero linearly in time.
 At any instant {\it before} the asymptotic value of the reflection
probability is attained, the time evolving reflection probability
in the region $-\infty <x\leq x\prime$   is given by
\begin{equation}
\label{3}
\left| R(t)\right|^{2}=\int ^{x\prime }_{-\infty }\left| \psi \left( x,t\right) \right|^{2}dx
\end{equation}
We denote the reflected probability for the static and the perturbed cases 
as $R_s(t)$ and $R_p(t)$ respectively. In \cite{bandyo} we computed these 
probabilities versus
time  for various values of $\epsilon$ which is
the time span over which the barrier height goes to zero (implying
different rates of reduction of the potential barrier).  We observed that 
$R_p(t) > R_s(t)$ during the time interval $t_d<t<t_c$. If
$t_{p}$  is the instant at which the perturbation starts,
$t_{c}$ the instant when the static and the perturbed  curves cross each other, and $t_{d}$ 
the time  from which the curve corresponding to the perturbed case starts
deviating from that in the unperturbed case,  we found that 
$t_{c}>t_{d}>t_{p}$.

Let us now consider the case when initially there is no barrier, and the
wave packet is allowed to propagate freely towards the right. A second 
detector placed far away at $x''$ records the time-dependent transmission probability
$T_s(t)$ (counting the transmitted particles up to various instants of time).
If a barrier is raised in the path of the wave packet, a portion of it will
be reflected back. We denote by $T_p(t)$ the transmitted probability in this
case. At any instant {\it before} the asymptotic value of the transmission
probability ($=1$ since there is no absorption) is attained, the time 
evolving transmission probability
in the region $x'' \leq x \leq \infty$   is given by
\begin{equation}
\label{3}
\left| T(t)\right|^{2}=\int _{x''}^{\infty }\left| \psi \left( x,t\right) \right|^{2}dx
\end{equation}
We compute the values of $T_s(t)$ and $T_p(t)$ using the same method of
numerically integrating the time dependent Schrodinger equation as used
in \cite{bandyo}, which was first developed in \cite{goldberg}.
The following values for the parameters are chosen for  our computations 
(in units of $\hbar=1$ and $m=1/2$): $x_0=-0.3$,
$\sigma=0.05/\sqrt{2}$, $x_c=0$, $w=0.016$, $x'=-0.5$, $x''=0.5$ and 
$t_p = 8\times 10^{-4}$. It should be emphasized that the observation of
the phenomenon of superarrivals does {\it not} hinge upon the choice
of these particular values of the parameters. Indeed, the quantitative
dependence of superarrivals on the parameter values have been studied
in \cite{bandyo} where it was shown that superarrivals in reflection
persist for a sufficiently wide range of values of these parameters.
We choose one particular set of values for the computations used in
this paper since our aim here is primarily to investigate the origin
of superarrivals.

The potential barrier is raised from $V=0$ to $V=2E$ (where $E$ is the
energy of the incident wave packet) linearly in time $\epsilon$.
In Figure 1 we plot the computed values of $T_s(t)$ and $T_p(t)$ for different
values of $\epsilon$. The numbers denoting various instants of time in
this as well as the subsequent figures are in units of the time steps used
in the numerical algorithm. For example, $t = 8\times 10^{-4}$ corresponds
to $400$ time steps. 
It is seen that superarrivals are also exhibited in the transmitted wave packet.

\begin{figure}
\begin{center}
\centerline{\epsfig{file=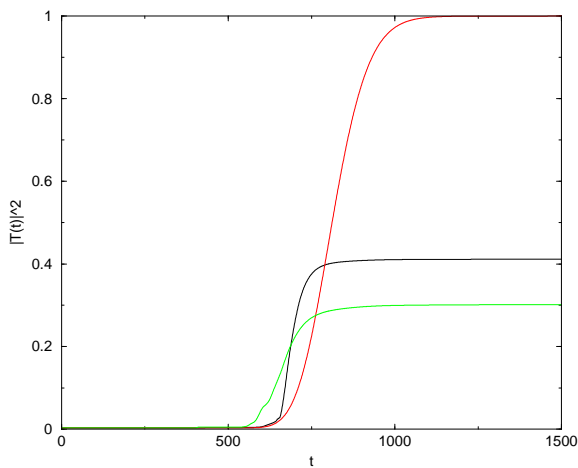,height=3.0in,angle=270}}
\narrowtext{\caption{The transmission probability $|T(t)|^2$ is plotted for 
various values of
$\epsilon$. The top curve  reaches value $1$ asymptotically and
corresponds to the zero barrier case. The next two curves with 
($\epsilon = 10$) and
 ($\epsilon = 40$) respectively, represent the transmission probabilities
for the rising barriers.}}
\end{center}
\end{figure}

Superarrivals can be quantitatively defined by a parameter $\eta$  given by
\begin{equation}
\label{8}
\eta =\frac{I_{p}-I_{s}}{I_{s}}
\end{equation}
where the quantities $I_{p}$  and $I_{s}$  are defined with respect to
$\Delta t = t_c-t_d$ during which superarrivals occur. For the case
of superarrivals in the reflected probability,
\begin{equation}
\label{9a}
I_{p}=\int _{\Delta t}\left| R_{p}(t)\right|^{2}dt
\end{equation}
\begin{equation}
\label{9b}
I_{s}=\int _{\Delta t}\left| R_{s}(t)\right|^{2}dt
\end{equation}
Replacing the static and perturbed reflected probabilities by $T_s(t)$ and 
$T_p(t)$ respectively, one can obtain the corresponding expression of $\eta$
for the case of the transmitted wave packet. 

It has been observed\cite{bandyo} that both
$\Delta t$  and superarrivals given by $\eta$  
depend on the instant $t_{p}$ around which the barrier is perturbed.
The magnitude of superarrivals
is appreciable only in cases where the wave packet has some significant 
overlap with the barrier while it is being perturbed. 
The magnitude of superarrivals falls off with increasing $\epsilon$,
for the reflected as well as the transmitted wave packets. 
Another interesting observation is about information transfer from
the perturbing barrier to the detector. We defined signal velocity
\begin{equation}
v_{e}=\frac{D}{t_{d}-(t_{p}-\frac{\epsilon }{2})}
\end{equation}
measuring how fast the influence of barrier perturbation travels
across the wave packet. We found that $v_e$ is again proportional to
$\epsilon$ as was the case with $\eta$. These features lead one to
argue that the wave packet acts as a carrier (objective field-like
behaviour) through which
information about the barrier perturbation propagates with a velocity that
is proportional to the ``disturbance'' (measured in terms of the rate of 
barrier reduction) imparted to the packet by the barrier.

Now, in order to  understand {\it how} superarrivals originate, we use
the concept of {\it particle trajectories} in terms of the Bohm
model (BM). We recall that
 BM provides an  ontological and a
self-consistent interpretation of the formalism of
quantum mechanics\cite{holland,squires}.
Predictions of BM
 are in agreement with that of standard quantum mechanics. In  BM
a wave function $\psi$ is  taken
to be an incomplete specification of the state of an individual
particle. An objectively real ``position'' coordinate
(``position'' existing irrespective of any external
observation) is
ascribed to a  particle apart from the wave function. Its  ``position''
evolves with time obeying an equation that can be
justified in the following way from the Schroedinger equation (considering
the one dimensional case)\cite{squires}
\begin{eqnarray}
i\hbar {\partial\psi \over \partial t} = H\psi \equiv - {\hbar^2
\over 2m} {\partial^2 \psi \over \partial x^2} + V(x)\psi
\end{eqnarray}
by writing
\begin{eqnarray}
\psi = Re^{iS/\hbar}
\end{eqnarray}
and using the continuity equation
\begin{eqnarray}
{\partial \over \partial x} (\rho v) + {\partial\rho \over \partial
t} = 0
\end{eqnarray}
with the probability distribution $\rho(x,t)$ being given by
\begin{eqnarray}
\rho = \vert \psi \vert^2.
\end{eqnarray}
It is important to note that $\rho$ in BM is ascribed an {\it
ontological}
significance by regarding it as representing the probability
density of ``particles'' occupying {\it actual} positions and the velocity 
$v$  is interpreted as an ontological (premeasurement) velocity. On the other
hand, in the standard interpretation, $\rho$ is interpreted as the
probability density of {\it finding} particles around specific
positions and there is no concept of an ontological velocity. Integrating 
Eq.(9) by using Eqs.(7), (8) and (10) and requiring that $v$ should 
vanish when $\rho$
vanishes  leads  to the Bohmian equation of motion where the particle
velocity $v(x,t)$ is given by
\begin{eqnarray}
v \equiv {dx \over dt} = {1\over m}{\partial S \over \partial x}
\end{eqnarray}
The particle trajectory
is thus deterministic  and is obtained by integrating the velocity equation
for a given initial position.

Another perspective on the notion of particle trajectories in BM is obtained by
decomposing the Schrodinger equation in terms of two real equations for the 
modulus $R$ and the phase $S$ of the wave function $\psi$\cite{holland}
\begin{eqnarray}
{\partial S \over \partial t} + {(\vec{\nabla} S)^2 \over 2m} - {\hbar^2 \over 2m}
{\nabla^2 R \over R} + V = 0 \\
{\partial R^2 \over \partial t} + \vec{\nabla}.\biggl({R^2 \vec{\nabla} S \over m}\biggr) = 0
\end{eqnarray}
and by indentifying
\begin{equation}
Q(x,t) = -{\hbar ^2 \over 2m}{\nabla^2 R \over R}
\end{equation}
as the ``quantum potential''\cite{holland}. The equation of motion of a particle
along its trajectory can now be written in a form analogous to Newton's second
law
\begin{equation}
{d \over dt}(m\dot{\vec{X}}) = - \vec{\nabla}(V + Q)|_{X}
\end{equation}
(with $d/dt = \partial/\partial t + \dot{\vec{X}}.\vec{\nabla}$) where the
particle is subjected to a quantum force $-\vec{\nabla} Q$ in addition to the
classical force $-\vec{\nabla}V$. The effective potential on the particle is
$(Q+V)$. We plot the profile of $Q$ versus $x$ at various instants 
of time near the 
potential barrier (when its height is reduced) in Figure 2. It is then transparent how the perturbation of the classical potential $V$ affects $Q$ away from the 
vicinity of the boundary of $V$. This in turn accounts for the sharp turn experienced
by those particles which contribute towards {\it superarrivals} (as we shall see explicitly later).

\begin{figure}
\begin{center}
\centerline{\epsfig{file=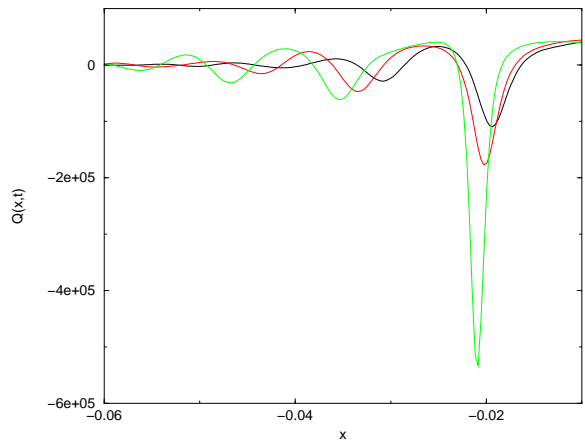,height=3.0in,angle=270}}
\narrowtext{\caption{Snapshots of the quantum potential $Q(x,t)$ are plotted 
versus $x$ at various instants of time. The potential barrier is located
in the region $-0.008 < x < 0.008$. Barrier perturbation is 
from $t=400$ to $t=410$.  The full, dashed and dotted 
curves represent $Q$ at times $t=420$, $425$ and $430$ respectively.
The wells in the quantum potential move towards the left with time
and reflect incoming particles away from the vicinity of the classical
barrier. This explains why certain particles arrive at the detector earlier
than they would have done if reflected from a static barrier. }}
\end{center}
\end{figure}

We compute the Bohmian trajectories for a given set of initial
positions with a Gaussian distribution corresponding to  the 
initial wave packet. This procedure is carried out for both the cases
of lowering and raising the barrier.  Since our purpose 
is to obtain conceptual clarity of the phenomenon of superarrivals,
it suffices to illustrate our scheme through the example of superarrivals
in the reflection probability when the barrier is reduced. 
All the qualitative as well as quantitative features of superarrivals
are similar in the case where one observes the transmitted probabilty
from a rising barrier. Thus, henceforth we consider only the
former case in the following discussion.
 
The following approach is used to study {\it superarrivals}
in terms of the Bohmian trajectories. First, a particular value of the barrier 
reduction rate, or $\epsilon$ is chosen.
We then choose a range of initial positions for which the trajectory arrival times at
the detector lie between $t_d$ and $t_c$ (i.e., we select only those trajectories which
{\it contribute} to superarrivals). We consider $N$ such 
trajectories whose initial positions form a Gaussian distribution. Let us denote {\it one} such
trajectory by $S_{ip}$ having the initial position $x_{i}$ and the arrival time $t_{ip}$.
Taking the static case,  the trajectory $S_i$
for that initial position $x_i$ is computed. Let the corresponding arrival time be $t_i$.
A supearrival parameter $\beta_i$ for the $i$-th Bohmian
trajectory is then defined as
\begin{equation}
\beta_i = {t_i - t_{ip} \over t_i}
\end{equation}
which provides a measure of superarrivals for a {\it particular value} 
of initial position. Next we define an {\it average value} 
\begin{equation}
\tilde{\beta} = {\sum_i \beta_i \over N}
\end{equation}
which provides a {\it quantitative estimate} of superarrivals obtained through 
Bohmian trajectories. 
%\vskip 0.2in

Our results  show that the arrival time\footnote{For conceptual subtleties 
concerning {\it arrival time} in
the Bohm model, and, in general, its definition in quantum theory, see
\cite{arrtim}.} $t_{ip}$ for
the perturbed case is sensitive to the value of initial
position $x_{i}$. We have checked that for a particular initial position, $t_i$
 {\it exceeds} $t_{ip}$ for {\it only}  those trajectories which {\it contribute}
to superarrivals. This is a distinct feature associated with the
superarrivals  that can be identified in terms of the
Bohmian trajectories.
We plot a set of Bohmian trajectories in Figure 3. Note that the trajectories
of the particles corresponding to the perturbed case take a sharp turn and
arrive at the detector {\it earlier} than they would have for a static barrier. 
Any abrupt perturbation of the potential barrier has thus a {\it global  effect} on the
wave function  and affects the values of the quantum potential $Q(x,t)$ at various
points. Then, through the Bohmian equation of motion the 
velocities of the incident particles get correspondingly affected 
{\it much before}  reaching the vicinity of the potential
barrier. Superarrivals originate from {\it those} particles in the
perturbed case which reach the detector {\it earlier} than those
corresponding to the {\it same} initial positions in the static
case. This accounts for why a detector records more counts in the perturbed case during 
a particular time interval as compared to that in a 
static situation. The origin of superarrivals can thus be
understood in this way by using the Bohmian trajectories. 

\begin{figure}
\begin{center}
\centerline{\epsfig{file=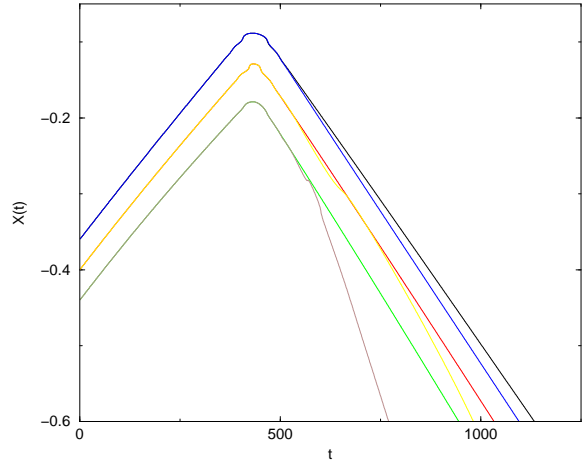,height=3.0in,angle=270}}
\narrowtext{\caption{Bohmian trajectories for particles originating from 
the same initial positions get reflected differently from the static
and the perturbed barriers. The trajectories undergo sharper turns
when the barrier is perturbed and arrive the the detector earlier
than they would have done for the static barrier case.
 The barrier is placed at $x=-0.008$ to $x=0.008$. 
Perturbation takes place from $t=400$ to $t=410$. }}
\end{center}
\end{figure}

The effect of altering the barrier perturbation time $\epsilon$ on the magnitude
of superarrivals $\tilde{\beta}$ can be studied by computing $\tilde{\beta}$ 
for various values of $\epsilon$. We display the results of this study in Figure 4.
Note that the the magnitude of superarrivals decreases monotonically with
increasing  $\epsilon$, or decreasing rate of perturbation. This effect was
also observed in \cite{bandyo} where we obtained a similar behaviour for
the superarrival parameter $\eta$. The similarity of these two results 
obtained through entirely different techniques reinforces our contention
about the dynamical nature of superarrivals originating from a ``disturbance'' 
provided by the lowering of potential barrier, which propagates across the wave 
function with a definite speed.

\begin{figure}
\begin{center}
\centerline{\epsfig{file=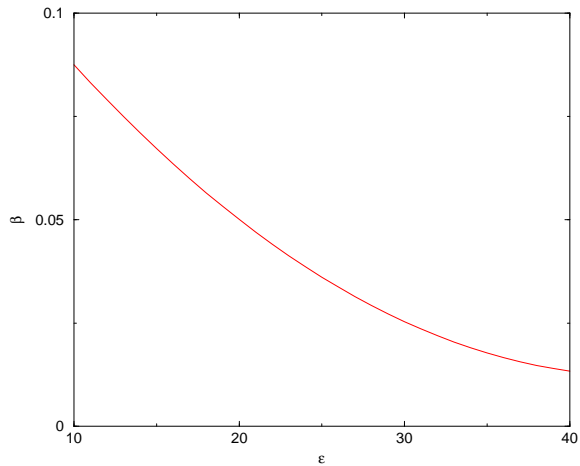,height=3.0in,angle=270}}
\narrowtext{\caption{The Bohmian superarrival parameter $\tilde{\beta}$ is
plotted versus $\epsilon$. }}
\end{center}
\end{figure}

To conclude, in this paper we have explored further the nature and origin
of {\it quantum superarrivals} manifested in terms of enhanced reflection
and transmission probabilities of  wave packets from a perturbed potential 
barrier which is respectively lowered or raised.
We have shown how the concept of particle trajectories
obtained from the Bohm model enables one to have an insight into the
phenomenon of quantum superarrivals. This analysis substantiates 
our earlier
contention\cite{bandyo} that superarrivals arise from a dynamical 
disturbance provided by the perturbed barrier which 
propagates across the wave packet (which acts like a ``physically real field'')
with a definite speed and affects the ``particles''.
Such time dependent quantum phenomena could be useful in furnishing
examples of the {\it conceptual utility} of the Bohm model from a perspective different from other 
 examples studied recently \cite{pla} for this purpose.

\vskip 0.1in 
D. Home thanks John Corbett for useful suggestions.
Md. Manirul Ali acknowledges the grant of a research fellowship from
Council of Scientific and Industrial Research, India.

\end{multicols}


\begin{thebibliography}{99}

\bibitem{greenberger}
D. M. Greenberger, Physica B {\bf 151}, 374 (1988);
M.V. Berry, J. Phys. A {\bf 29}, 6617 (1996); M.V. Berry and S. Klein,
J. Mod. Opt. {\bf 43}, 2139 (1996); 
 D.L. Aronstein and C.R. Stroud, Phys. Rev. A {\bf 55}, 4526 (1997);
 F. Lillo and R. N. Mantegna, Phys. Rev. Lett. {\bf 84}, 1061 (2000);    
 G. Kalbermann, J. Phys. A {\bf 34}, 3841 (2001); G. Kalbermann, 
quant-ph/0203036.
\bibitem{venu}
See, for instance,
 A. Venugopalan and G.S. Agarwal, Phys. Rev. A {\bf 59}, 1413 (1999);
F.B.J. Buchkremer, R. Dumke, H. Levsen, G. Birkl and W. Ertmer, 
Phys. Rev. Lett. {\bf 85}, 3121 (2000); H. Mack, M. Bienert, F. Haug, F. S. 
Straub, M. Freyberger and W. P. Schleich, quant-ph/0204040.  
\bibitem{bandyo}
S. Bandyopadhyay, A. S. Majumdar and D. Home, Phys. Rev. A {\bf 65}, 052718 
(2002). 
\bibitem{goldberg}
A. Goldberg, H. M. Schey and J. L. Schwartz, Am. J. Phys. {\bf 35}, 177 (1967).
\bibitem{bohm} 
D.Bohm, Phys. Rev. {\bf 85}, 166 (1952); D.Bohm and
B.J.Hiley, ``The Undivided Universe'', (Routledge, London, 1993).
\bibitem{holland}
 P.R.Holland, ``The Quantum Theory of Motion'', (Cambridge
University Press, London, 1993).
\bibitem{squires}
E.J. Squires, in ``Bohmian Mechanics and Quantum Theory: An Appraisal'',
Eds. J.T. Cushing, A. Fine and S. Goldstein (Kluwer, Dordrecht, 1996),
pp. 131-140. 
\bibitem{arrtim}
J. G. Muga and C. R. Leavens, Phys. Rep. {\bf 338}, 353 (2000), and 
references therein; G. Gruebl and K. Rheinberger, quant-ph/0202084. 
\bibitem{pla}
P. Ghose, A. S. Majumdar, S. Guha and J. Sau, Phys. Lett. A {\bf 290}, 
205 (2001); A. S. Majumdar and D. Home, Phys. Lett. A {\bf 296}, 176 (2002).
\end{thebibliography}
\end{document}